\documentstyle [aps,pre,epsf]{revtex}
\newcommand{\half}   {\frac{1}{2}}
\newcommand{\asix}   {\frac{a^2}{6}}
\newcommand{\rr   }   { {\bf r} }
\newcommand{\rls }   { {\bf r}_l(s)}
\newcommand{\rlsa}   { {\bf r}^\alpha_l(s)}
\newcommand{\rld }   { \dot{\bf r}_l}
\newcommand{\rlda}   { \dot{\bf r}^\alpha_l}
\newcommand{\Rp  }   { {\bf R}_i^+ }
\newcommand{\Rm  }   { {\bf R}_j^- }
\newcommand{\q   }   { {\bf q}    }
\newcommand{\dr  }   { {\rm d}\rr\ }
\newcommand{\dra }   { {\rm d}\rr^\alpha\ }
\newcommand{\drr }   { {\rm d}\rr'\ }
\newcommand{\dq  }   {{{\rm d}{\bf q}\over (2\pi)^3}\ }
\newcommand{\ds  }   { {\rm d}s\ }

\newcommand{\D   }[1]{ {\cal D}#1\ }

\newcommand{\eps }   {\varepsilon}

\newcommand{\e   }   { {\rm e} }
\newcommand{\kt  }   {k_BT}

\begin{document}
\draft

\title {\bf Random Polyelectrolytes and Polyampholytes in Solution}

\author {Itamar Borukhov and David Andelman}
\address{School of Physics and Astronomy,
         Raymond and Beverly Sackler
         Faculty of Exact Sciences\\
         Tel-Aviv University,
         Ramat-Aviv 69978, Tel-Aviv, Israel}

\author {Henri Orland}
\address{Service de Physique Th\' eorique,
         CE-Saclay,
         91191 Gif-sur-Yvette Cedex, France }

\date{\today}

\maketitle

\begin{abstract}
   The behavior of polyelectrolytes and polyampholytes
in semi--dilute solutions is investigated theoretically.
    Various  statistical 
charge distributions along the polyelectrolyte chains 
are considered: 
(i) {\em smeared}, where  the charges are uniformly distributed
along the chain. 
(ii) {\em Annealed}, where the charges are allowed to 
associate and dissociate from the chain.
(iii) {\em Permuted}, where the total number of charges on the 
chain is fixed, but the charges can move along the chain.
(iv) {\em Quenched}, where the charges on the chains 
are ``frozen'' in a random configuration.
Finally, we also consider (v) {\em polyampholytes}, 
where each  monomer can be positively or negatively charged,
 or neutral.
Path integral formulation was used to
 derive mean field free energies for the different models.
Self-consistent field equation is obtained for the 
polymer order parameter and 
a Poisson--Boltzmann like equation for the 
electrostatic potential.
  We show that the difference between the permuted 
and the smeared models 
is a constant shift in the chemical potential 
leading to similar  mean field equations.
 Within mean--field the quenched model is found to be equivalent 
to the annealed one, provided that
 the system is coupled to a reservoir of polyelectrolyte
chains.
    The random phase approximation is used 
to calculate  the monomer--monomer structure factor $S(q)$
for the different statistical charge distribution models.
We show that in the annealed model, 
fluctuations of the 
the monomer charges 
contribute to the electrostatic 
screening in addition to the 
free ions in the solution.
The strength of this screening depends on the variance of the 
monomer charge distribution and is especially important
for polyampholytes in bad solvent conditions where the
mesophase separation is enhanced.
The ratio between the variance and the
net average charge determines whether  polyampholytes
behave as polyelectrolytes or as neutral chains. 
\end{abstract}



\pagestyle{plain}
\section{Introduction}

  Charged polymers have drawn a considerable amount
of attention 
in the past years both theoretically and experimentally
 \cite{Oosawa,deGennesBook,BarratJoanny}.
This is due, on one hand, 
to their wide range of industrial applications
 in processes involving charged colloids 
\cite{Cabane,Dickinson}
and on the other hand, to their resemblance to 
water soluble bio-polymers
such as proteins and DNA.
  A distinction is made between {\em polyelectrolytes}, with 
all charges having the same sign, 
and {\em polyampholytes} with positively 
as well as negatively charged monomers
\cite{Wittmer,Kantor,Dobrynin_PA,Everaers,Thirumalai}.
For the former, the electrostatic interactions are 
repulsive and long ranged.
 For the latter, the repulsion of like--charges 
competes with the attraction of opposite charges, 
resulting in a complex behavior which depends on the 
net charge of the chain.

  In spite of extensive investigations of polyelectrolytes 
including the pioneering works of Katchalsky et al.
\cite{Katchalsky} and Manning \cite{Manning}, 
polyelectrolytes are much less understood than neutral
polymers.
For example, there  exists a 
debate regarding the persistence length
and the chain conformations for
single chains in dilute solutions
\cite{Odijk77,Skolnick,Witten,deGennes76,Pfeuty,Kremer}. 
This is due to the delicate interplay between the 
chain connectivity and the long range 
nature of electrostatic interactions.

   Semi--dilute polyelectrolyte 
solutions where the chains interact 
with one another have been studied as well 
\cite{deGennes76,Pfeuty,Odijk79,Dobrynin}.
In the so-called ``blob'' picture, 
scaling laws are derived by 
 singling
out the most dominant
interaction at different length scales.
Another technique is the Random Phase Approximation (RPA) 
\cite{Erukhimovich,LeiblerJoanny,Raphael,Vilgis,Dormidontova,Liu} 
 used for the calculation 
of the structure factor $S(q)$ as can be measured by 
scattering experiments \cite{Candau,Schmidt}.

In addition to bulk properties, 
special attention was directed to adsorption experiments 
of polyelectrolytes onto a single charged surface 
\cite{CS-1988,CS-chapter,FleerBook,Norde},
and between two charged surfaces 
\cite{Klein,Marra,Claesson,Bergeron}.
On the theoretical side, 
 discrete models have been 
employed for which the chains are placed on a lattice 
\cite{vanderSchee,Papenhuijzen,Evers,vandeSteegSim}. This approach 
has been used by B\"ohmer et al. 
\cite{Bohmer} to calculate force curves 
between two charged surfaces containing a polyelectrolyte solution.

  Another approach is a continuum one
\cite{Muthukumar,Varoqui91,Varoqui93,epl,macromols} where
the charge densities, monomer densities and electric field are
treated as continuous functions of the local position.
Varoqui et al. \cite{Varoqui91,Varoqui93} 
investigated polyelectrolyte adsorption onto one surface, 
while Podgornik \cite{Podgornik}   
calculated inter--surface forces. In  recent
works \cite{macromols} 
non-linear excluded volume interactions
have been  included and 
 scaling laws  characterizing the adsorption 
of polyelectrolytes are proposed
 (within mean-field approximation).

   In the present work the continuum approach
 is extended to {\it random} (heterogeneous) polyelectrolytes.
We study
several statistical charge distributions 
(i.e. annealed, quenched and permuted) corresponding
to   different
physical situations.
Polyampholytes with positive and negative  charges
are studied as well.
One of the goals of the present work is to
take explicitly into account
several  characteristics of polyelectrolytes
such as
 the connectivity of the polymer chains, 
the non-electrostatic monomer-monomer interactions
and the Coulomb interactions between 
charged monomers, 
counter-ions and co-ions.

   The paper is organized as follows:
   in the next section we present a
general formalism for calculating the 
free energy of randomly  charged chains.
This formalism is applied in Sec.~III  to
derive mean-field equations for the various 
charge distributions (including polyampholytes).  
The reader who is not interested in the  technical details
can skip these two sections and go directly to Sec.~IV where
the mean-field results are summarized
and the various charge distributions are compared.
 The random phase approximation (RPA) is used in Sec.~V to 
calculate the structure factor $S(q)$ of polyelectrolytes and 
polyampholytes  in good and bad solvents.

\section{General Formalism}

   Let us consider a semi--dilute solution of polyelectrolytes 
in a good solvent in  presence of salt (electrolyte). 
The system
is schematically drawn on Fig.~1.
In  the model, based on the Edwards' formalism \cite{Edwards65}, 
the microscopic degrees of freedom are  
the monomer positions $\{\rls\}$, where $s\in[0,N]$ 
is a continuous index along a chain of $N$ monomers and 
 $l=1,...,M$ is the label of the $M$ chains in solution.
The positions of the small co-ions (counter-ions) are 
denoted by $\Rp$ ($\Rm$) where $i=1,...,N^+$ and 
 $j=1,...,N^-$ are, respectively, the indices of the monovalent
positive ($+e$) and negative ($-e$) ions (see Fig.~1).
For simplicity only a symmetric 1:1 electrolyte is considered.

   The partition function $Z$ is then expressed as a path integral 
over all possible configurations: 
\begin{eqnarray}
   Z &=& \int \D{\rls} \D{\Rp} \D{\Rm}
   \exp\left(
       - \frac{3}{2a^2} \sum_{l=1}^M \int_0^N \ds \rld^2(s)
   \right) 
    \nonumber \\ &\times&
   \exp\left(
        - \half\beta\int\dr\drr \hat\rho_{\rm c} (\rr)v_{\rm c}(\rr-\rr')
                                \hat\rho_{\rm c} (\rr')
       - \half v   \int \dr \hat\rho_{\rm m}^2  
    \right)
   \label{defZ}
\end{eqnarray}
where $\int\D{g}$ 
denotes the functional integral over the function $g(\rr)$.
   The first term in the exponent is the Wiener measure 
representing the connectivity of the polymer random walk
\cite{Wiegel}, where $\rld(s)$ is the derivative of $\rls$ 
with respect to the monomer index $s$ 
and the Kuhn length $a$ is the effective 
monomer length. In writing this term, we 
have assumed that the charged chains
can be modeled as flexible chains. 
This assumption 
can be more easily
justified for semi-dilute solutions
of weakly charged chains.

   The second term is the electrostatic interaction term where
 $\beta=1/k_BT$ is the inverse thermal energy, 
 $v_{\rm c}(\rr)= 1/\eps|\rr|$ is the Coulomb interaction,
 $\eps $ is the dielectric constant of the solution 
and $\hat\rho_{\rm c} (\rr)$ is the local {\it charge} density operator
including all charges in the solution 
(charged monomers, co-ions and counter-ions):
\begin{eqnarray}
   \hat\rho_{\rm c} (\rr) = 
       \sum_{l=1}^M \int_0^N \ds q_l(s)\delta(\rr-\rls)
      +\sum_{i=1}^{N^+} e\delta(\rr-\Rp)
      -\sum_{j=1}^{N^-} e\delta(\rr-\Rm)
   \label{def_rhoc}
\end{eqnarray}
  where $q_l(s)$ is the 
random variable denoting the charge carried by
the $s$ monomer along the $l$ chain.
In the next section we will consider several 
charge distributions for $q_l(s)$. 
For example, in the smeared model, $q_l(s)=p e$ is a constant
independent of the position $s$ along the chain and the chain 
index $l$.

   The last term in eq.~\ref{defZ} is 
the excluded volume repulsion between monomers, where 
 $v\sim a^3$ is the excluded volume parameter and 
\begin{eqnarray}
   \hat\rho_{\rm m}(\rr) = \sum_{l=1}^M \int_0^N \ds \delta(\rr-\rls)
\label{def_rhom}
\end{eqnarray}
is the local {\it monomer} concentration operator.

It is possible to integrate out the microscopic
degrees of freedom in the partition function eq.~\ref{defZ} by
introducing
two pairs of collective coordinates: 
(i) the local monomer concentration $\rho_{\rm m}(\rr)$ and its 
conjugate field $\varphi_{\rm m}(\rr)$; and, 
(ii) the local charge density $\rho_{\rm c} (\rr)$ and its 
conjugate field $\varphi_{\rm c}(\rr)$.
This is done using the following identity
\begin{eqnarray}
  1 
    = \int\D{\rho_{\rm m}}\delta(\rho_{\rm m} - \hat\rho_{\rm m})
    = \int\D{\rho_{\rm m}}\D{\varphi_{\rm m}} \exp\left(
          i v\int\dr\rho_{\rm m}\varphi_{\rm m}
         -i v\sum_{l=1}^M \int_0^N \ds \varphi_{\rm m}[\rls]
      \right)
  \label{collective} 
\end{eqnarray}
   for $\rho_{\rm m}(\rr)$ and an analogous identity for $\rho_{\rm c} 
(\rr)$.
We will see below that $i\varphi_{\rm c}(\rr)$ is in fact the 
electrostatic potential.
The functional integrals over $\rho_{\rm m}(\rr)$ and 
$\rho_{\rm c}(\rr)$ 
are Gaussian integrals and are evaluated exactly, leading to
\begin{eqnarray}
  Z = \int \D{\varphi_{\rm m}}\D{\varphi_{\rm c}}
      \exp\left(- \int\dr \left\{ 
          \beta\frac{\eps}{8\pi}|\nabla\varphi_{\rm c}|^2
          +\half v\varphi_{\rm m}^2 \right\} \right)
      \  \zeta_+ \  \zeta_- \  \zeta_{\rm p}
  \label{Zvarphi}
\end{eqnarray}
   where 
 $\zeta_\pm=(\int\dr \e^{\mp i\beta e\varphi_{\rm c}(\rr)} )^{N^\pm}$ 
and $\zeta_{\rm p}$
are, respectively, the partition functions of the (small) co-ions,  counter-ions and 
polymer in the presence of the external fields $i\varphi_{\rm c}(\rr)$ and
$i\varphi_{\rm m}(\rr)$.
In the thermodynamic limit, where $N^\pm$ and the volume $V$ 
become large while the bulk concentrations $c_{\rm b}^\pm=N^\pm/V$
remain fixed,
 $\zeta_\pm$  become (up to a normalization constant):
\begin{eqnarray} 
  \zeta_\pm = \left[1 + {1\over V}\int\dr 
                 \left(\e^{\mp i\beta e\varphi_{\rm c}(\rr)}-1\right) 
              \right]^{Vc_{\rm b}^\pm}
       \ \longrightarrow\ 
  \exp\left( c_{\rm b}^\pm \int\dr 
     \left\{ \e^{\mp i\beta e\varphi_{\rm c}(\rr)}-1 \right\}\right)
  \label{zetapm}
\end{eqnarray}

   The partition function of the 
polymer chains $\zeta_{\rm p}$ 
in the presence of the two external fields is:
\begin{eqnarray}
  \zeta_{\rm p} = 
     \int\D{\rls} \exp\left(
       - \sum_{l=1}^M \int_0^N \ds \left\{ 
            \frac{3}{2a^2} \rld^2(s)
          + iv\varphi_{\rm m}[\rls]
          + i\beta q_l(s)\varphi_{\rm c}[\rls] \right\} 
     \right)
  \label{zetap}
\end{eqnarray}
   Note that the calculation of $\zeta_{\rm p}$ 
depends on the specific charge distribution for $q_l(s)$.

\section{Mean--Field Equations}
We apply the general formalism introduced above to different
monomer charge distributions: smeared, annealed, permuted, quenched
and  polyampholytes.
These charge distributions are applicable to different experimental systems.

\subsection{``Smeared" Polyelectrolytes}
The simplest model of a charge distribution is the so-called
    {\it smeared} polyelectrolyte. For a polyelectrolyte with 
    a fraction $p$ of its monomers being charged, this model
    assumes that
    each monomer carries a uniform fractional charge $pe $,
    where $e$ is the electron charge. Namely, $q_l(s)=pe$
    for any monomer $s$ on any chain $l$.
    Without loss of generality we shall assume
    that the polymer is positively charged.

 It is
possible here to use a well-known analogy from
 quantum mechanics to calculate 
the path integral of eq.~\ref{zetap}.
The partition function is analogous to a Feynman integral of the Hamiltonian 
 ${\cal H} = -\asix\nabla^2 + iv\varphi_{\rm m} 
+ i\beta p e\varphi_{\rm c}$
with imaginary time $t\rightarrow is$. 
Thus, each eigenstate $\phi_\nu$ 
contributes a term of the form $\exp(-MN E_\nu)$
where $MN$ is the total number of monomers in the solution.
In the thermodynamic limit $N\rightarrow\infty $,
the ground state dominates
over all other eigenstates  \cite{vanOpheusden},
and $\zeta_{\rm p}$ reduces to:
\begin{eqnarray}
  \zeta_{\rm p} \approx \e^{-M N E_0} = 
          \exp\left( - \int\dr \left\{
            \asix|\nabla\phi|^2 + iv\varphi_{\rm m}(\rr)\phi^2(\rr)
          + i \beta p e\varphi_{\rm c}(\rr)\phi^2(\rr) 
          -\mu_{\rm p}\phi^2(\rr) 
          \right\} \right)
  \label{zetas}
\end{eqnarray} 
    where $E_0$ is the ground-state energy,
 $\phi(\rr)$ is the renormalized ground-state eigenfunction 
and $\mu_{\rm p}$ is a Lagrange multiplier added in order to ensure
the normalization of the wave-function 
 $\phi_{\rm b}^2={1\over V}\int\dr \phi^2(\rr)$, 
 $\phi_{\rm b}^2$ being the bulk monomer concentration.
The polymer analog of the wave-function $\phi(\rr)$ 
is usually referred to as the {\em polymer order parameter}, 
and the local monomer concentration per unit volume
can be shown to be $\rho_{\rm m}(\rr)=\phi^2(\rr)$ \cite{deGennesBook}.

   The field $\varphi_{\rm m}$ can now be integrated out leaving
a functional integral only over $\psi\equiv i\varphi_{\rm c}$,
and the partition function reduces to: 
\begin{eqnarray}
  Z = \int \D{\psi} \exp\left( -\beta F_{\rm s} \right)
  \label{Zpsis}
\end{eqnarray}
   where
\begin{eqnarray}
  \beta F_{\rm s} = \int\dr \Biggl\{
    & - & \beta\frac{\eps}{8\pi}|\nabla\psi|^2
     + \sum_\pm c_{\rm b}^\pm (1-\e^{\mp \beta e\psi(\rr)}) 
  \nonumber\\
     &+& \asix|\nabla\phi|^2 + \half v\phi^4(\rr)
     -\mu_{\rm p}\phi^2(\rr)
     +  \beta p e\psi(\rr)\phi^2(\rr)
     \Biggr\}
  \label{betaFs}
\end{eqnarray}
Note that $\psi(\rr)$ 
is identified as the electrostatic potential.
   
Within a mean--field approximation, the functional integral 
is dominated by the saddle point given by the condition 
 $\delta{F}/\delta{\psi}=0$.
This results in a  Poisson--Boltzmann (PB) like
equation for the electrostatic potential,
which includes all charge sources
\begin{eqnarray}
   \nabla^2\psi(\rr)
   &=& \frac{8\pi e}{\varepsilon}c_{\rm b} \sinh(\beta e\psi)
    -  \frac{4\pi e}{\varepsilon}
       \left( p\phi^2 - p\phi_{\rm b}^2 \e^{\beta e\psi} \right)
   \label{PBs}
\end{eqnarray}
   The right hand side represents the local charge density.
The first term is the symmetric 1:1 electrolyte contribution, 
the second term is associated with the (positive) charges 
on the polymer chains,
while the last term comes form the counter-ions which 
dissociated from the chains. Note that  charge neutrality implies 
 $c_{\rm b}^{+}\rightarrow c_{\rm b}$ and 
 $c_{\rm b}^{-}\rightarrow c_{\rm b}+p\phi_{\rm b}^2$.

   Since $\phi(\rr) $ is the ground state eigenfunction, 
it satisfies the variational equation $\delta{F}/{\delta\phi}=0$, 
yielding a self-consistent field (SCF) equation for the 
polymer order parameter: 
\begin{eqnarray}
   \asix \nabla^2\phi(\rr) 
   &=& v(\phi^3-\phi_{\rm b}^2\phi) + \beta p e\psi\phi 
   \label{SCFs}
\end{eqnarray}
This is an equation for the density of polymer chains
in an external electric potential $\psi$ and with excluded
volume interactions characterized by the parameter $v$.
In the above equation
 $\mu_{\rm p}$ has been substituted in order to have the correct
bulk limit: $\psi \rightarrow 0$ and $\phi^2\rightarrow\phi_{\rm b}^2$.



\subsection{Annealed Polyelectrolytes}

The derivation presented above for the smeared case 
can be extended to more realistic annealed 
charge distributions where charges can dynamically
associate and dissociate from the chains.
 The annealed model describes an experimental system
    where the monomers have weak acidic (or basic) groups.
    The pH of the solution controls the 
    degree of association/dissociation of ions on the 
    polymer chain. 

Assuming no charge correlations along the chain,
the  monomer charge distribution $f[q_l(s)]$
is defined as the nominal probability of 
the monomer $s$ along the chain $l$ to have
a charge $q_l(s)$.
\begin{equation}
 f\bigl[q_l(s)\bigr]=\sum_{j}p_j\,\delta\bigl(q_l(s)-z_j e\bigr)
\label{fq}
\end{equation}
where the randomly charged 
chain is described by a set of valencies $\{z_j\}$
 $(z_j=0,  1,  2, \dots )$ with normalized probabilities $\{p_j\}$
such that $\sum_j p_j=1$. 
Here, we  concentrate on a simple
example of a polyelectrolyte  
for which each monomer can 
be either positively charged  
($z=1$ and $q=+e$) with probability $p$ or 
neutral with probability $1-p$.
 The charge distribution for each monomer is 
\begin{equation}
 f\bigl[q_l(s)\bigr] = p\delta\bigl(q_l(s)-e\bigr) 
                  + (1-p)\delta\bigl(q_l(s)\bigr)
\label{fqa}
\end{equation}
It is important to note that for {\it annealed} polyelectrolytes,
the partition function has to be averaged with respect to
the monomer charges since they are in thermal equilibrium
with the reservoir.
As a result the electrostatic contribution of one monomer 
in $\zeta_{\rm p}$ (eq.~\ref{zetap}) becomes:
\begin{eqnarray}
  \left\langle \e^{-i\beta q_l(s)\varphi_{\rm c}[\rls] } 
  \right\rangle_p
   = 1-p + p\e^{-i\beta e\varphi_{\rm c}[\rls] }
  \label{trqlsa}
\end{eqnarray}
   where the average is taken over the charge distribution,
 $\langle{\cal O}\rangle_p=\int {\cal O}(q)f(q)\,{\rm d}q$,
   and the annealed free energy becomes 
\begin{eqnarray}
  \beta F_{\rm a} = \int\dr \Biggl\{
     &-& \beta\frac{\eps}{8\pi}|\nabla\psi|^2
     + \sum_\pm c_{\rm b}^\pm (1-\e^{\mp \beta e\psi(\rr)}) 
\nonumber\\  
    & +&~ \asix|\nabla\phi|^2 + \half v\phi^4(\rr)
     -\mu_{\rm p}\phi^2(\rr)
     -\phi^2(\rr) \log\left( 1-p + p\e^{-\beta e\psi(\rr)} 
  \right) 
     \Biggr\}
 \label{betaFa}
\end{eqnarray}

The variation of eq.~\ref{betaFa} with respect to 
 $\psi(\rr)$, leads to a  PB like equation
\begin{eqnarray}
   \nabla^2\psi(\rr)
   &=& \frac{8\pi e}{\varepsilon}c_{\rm b} \sinh(\beta e\psi)
    -  \frac{4\pi e}{\varepsilon}
       \left( p_{\rm a}\phi^2 -
         p\phi_{\rm b}^2 \e^{\beta e\psi} \right)
       \label{PBa}
\end{eqnarray}
   where 
\begin{eqnarray}
 p_{\rm a}(\rr) \equiv
        p\e^{-\beta e\psi(\rr)}/(1-p+p\e^{-\beta e\psi(\rr)})
\end{eqnarray}
can be interpreted as the {\em annealed} charge probability
and depends on the local electrostatic potential.
Similarly, the variation of eq.~\ref{betaFa} with respect to
 $\phi(\rr)$ gives a modified SCF
equation:
\begin{eqnarray}
   \label{SCFa}
   \asix \nabla^2\phi(\rr) &=& v(\phi^3-\phi_{\rm b}^2\phi) 
       + \phi\log\left( 1-p_{\rm a} + p_{\rm a}\e^{\beta e\psi}\right)
\end{eqnarray}
Equations \ref{PBa} and \ref{SCFa} are the annealed equations
for the charge distribution eq. \ref{fqa}, 
similar to eqs. \ref{PBs} and \ref{SCFs} for the smeared case.
 
\subsection{``Permuted" Polyelectrolytes}

Another variant of the annealed case is 
the permuted model for which 
a fixed number of charges $pN$ are free 
to move along each chain without dissociating from it. 
Thus, the total charge on the chain remains constant. 
This is introduced into the annealed
model by adding a constraint in eq.~\ref{zetap} 
in order to keep the total charge on each of the chain fixed:
\begin{eqnarray}
   \prod_l\delta\left(\int_0^N \ds q_l(s) - Npe\right) = 
   \int{\rm d}u_l\ \exp\left( i u_l\int_0^N \ds [q_l(s)-pe] \right)
   \label{deful}
\end{eqnarray}
where $u_l$ is a Lagrange multiplier related to the 
fixed charge constraint of the chain $l$.
Assuming ground state dominance, the polymer partition function 
becomes:
\begin{eqnarray}
  \zeta_{\rm p} = \Biggl[ \int{\rm d}u\ 
          \exp\Biggl( - \int\dr \Biggl\{&&
            \asix|\nabla\phi|^2 
          + iv\varphi_{\rm m}(\rr)\phi^2(\rr) + ipe u\phi^2(\rr)
\nonumber \\          
         &-& \mu_{\rm p}\varphi_{\rm m}^2(\rr) 
          - \phi^2(\rr)
          \log\left(1+\e^{-i p\beta e\varphi_{\rm c}(\rr)+ieu}\right)
          \Biggr\} \Biggr) \Biggr]^M
  \label{zetaperm}
\end{eqnarray}
where the index $l$ is dropped from the functional integral
over $u_l$ since 
the constraint is satisfied separately on each chain.
   Note that the fraction $p$ of charged monomers is introduced 
through the constraint (eq.~\ref{deful}).
In order to carry out the integration 
  over $u$, we use the  identity 
(similar to eq.~\ref{collective})
\begin{eqnarray}
  1 = \int\D{g(\rr)}\delta(g(\rr) - u)
    = \int\D{g(\rr)}\D{h(\rr)} \exp\left(
          i \int\dr h(\rr)g(\rr) 
         -i \int\dr h(\rr)u
      \right) 
  \label{zetau}
\end{eqnarray}
A sequence of saddle point approximations for 
 $g(\rr)$, $h(\rr)$ and $u$ leads to the
following mean--field free energy for the permuted case:
\begin{eqnarray}
  \beta F_{\rm p} = \beta F_{\rm s}
        + \int\dr \left\{ p\log p ~ + ~(1-p)\log(1-p) \right\} \phi^2(\rr)
  \label{betafelp}
\end{eqnarray}
   This correction term represents the translational entropy 
of the charges on the chain \cite{Raphael}. 
However, the last term being quadratic in $\phi $ and 
independent of $\psi $,  it
only shifts the chemical potential
 $\mu_p$  
without affecting the differential equations 
\ref{PBs} and \ref{SCFs}.

\subsection{Quenched Polyelectrolytes}

In the quenched model the charge distribution is frozen. 
    Experimentally, this  corresponds to heterogeneous copolymers 
    with a random sequence of charged and neutral monomers.
    The specific sequence of each copolymer is determined 
    during the polymerization stage and represents one 
    possible realization of the random distribution. Various
    physical quantities are then calculated by 
    averaging over this random distribution.
Instead of averaging
the partition function, 
one should average the free energy itself  over the random 
charge distribution.
A standard method in quenched systems is
 the {\em replica trick} \cite{Replica} based on 
the following identity:
\begin{eqnarray}
  \left \langle \log{Z} \right \rangle  
= \lim_{n\rightarrow0}{{\rm d}\left \langle {Z^n}\right \rangle \over {\rm d}n}
  \label{logZ}
\end{eqnarray}
where $\langle \dots \rangle$ 
indicates an average over the quenched disorder and
the $n$-th power of the partition function introduces 
 $n$ replicas of the system all having the same monomer 
charge distribution. Generalizing the above approach
we obtain:
\begin{eqnarray}
  \left \langle {Z^n} \right \rangle &=& 
     \int \D{\varphi_{\rm m}^\alpha}\D{\varphi_{\rm c}^\alpha}
        \exp \left( - \sum_{\alpha=1}^n\int\dr \left\{ 
            \beta\frac{\eps}{8\pi}|\nabla\varphi_{\rm c}^\alpha|^2
           +\sum_\pm c_{\rm b}^\pm(1-\e^{\mp i\beta e\varphi_{\rm c}^\alpha})
           +\half v\,(\varphi_{\rm m}^\alpha)^2 
        \right\} \right)
        \times  \left \langle {\zeta_n}\right \rangle 
  \label{Zvarphiq}
\end{eqnarray}
where
\begin{eqnarray}
  \left \langle {\zeta_n} \right \rangle &=& 
     \int\D{\rlsa} \exp\left( 
        - \sum_{\alpha=1}^n \sum_{l=1}^M \int_0^N \ds
            \left\{ \frac{3}{2a^2} ({\rlda}(s))^2
          + iv\varphi_{\rm m}^\alpha[\rlsa] 
           \right\} \right)
        \nonumber \\
      &\times& \exp\left(
        \sum_{l=1}^M \int_0^N \ds
        \log\left(1-p + 
             p\e^{-i\beta e \sum_\alpha \varphi_{\rm c}^\alpha[\rlsa]}
             \right)
        \right)     
  \label{defzetan}
\end{eqnarray}
and the superscript $\alpha=1,\dots,n$ is the label of the 
 replica $\alpha$.
A mean--field estimate of $\zeta_n$ can be obtained by using again 
the quantum--mechanical analogy. An additional complication is 
that the effective Hamiltonian here is a {\it many-body} one as
it couples different replicas:
\begin{eqnarray}
{\cal H}_n = 
   \sum_{\alpha=1}^n \left\{ 
    - \asix\nabla_\alpha^2 + iv \varphi_{\rm m}^\alpha 
   \right\} - 
      \log\left(1-p + p\e^{-i\beta e\sum_\alpha\varphi_{\rm c}^\alpha}
   \right)
  \label{Hn}
\end{eqnarray}
   Assuming no replica symmetry breaking, we
use a Hartree approximation to express the ground state 
of the many-body 
eigenfunction 
 $\phi_n(\{\rr^\alpha\}) = \prod_{\alpha=1}^n \phi(\rr^\alpha)$,
in terms of single-body eigenfunctions $\phi(\rr^\alpha)$:
\begin{eqnarray}
  \zeta_n \approx \e^{-M E_n} = 
          \exp\left( - n M\int\dr \left\{
            \asix|\nabla\phi|^2 + iv\varphi_{\rm m}\phi^2
          -\mu_{\rm p}\varphi_{\rm m}^2 
          \right\} \right)
  \nonumber \\
     \times \exp\left( M\int\dra \prod_{\alpha=1}^n
            \phi^2(\rr^\alpha)
            \log(1-p + p\e^{-i\beta e\sum_{\alpha}\varphi_{\rm c}(\rr^\alpha)})
         \right)
  \label{zetan}
\end{eqnarray}
   The path integrals over $\varphi_{\rm m}^\alpha(\rr)$ 
can be evaluated exactly, whereas
   the path integrals over $\varphi_{\rm c}^\alpha(\rr)$ 
can be approximated by their saddle point values.
Since we assumed no replica symmetry breaking,
 all saddle point functions are identical: 
 $i\varphi_{\rm c}^\alpha(\rr)\equiv\psi(\rr)$.
When we take into account the coupling with the 
polyelectrolyte reservoir, this equation 
reduces to the annealed PB  equation 
(eq.~\ref{PBa}).
   Similarly, the SCF equation reduces 
to eq.~\ref{SCFa}.

\subsection{Annealed Polyampholytes}

    Finally, the mean--field formalism can be generalized in a 
straightforward way to treat {polyampholytes}, 
consisting of negatively and positively charged monomers.
A general polyampholyte is described by a set of valencies
 $\{z_j\}$ ($z_j=0,\ \pm1,\ \pm2,\ \dots$) 
with probabilities $\{p_j\}$, 
where $\sum_j p_j=1$.
The statistical distribution of the charges can be either 
annealed, quenched or permuted.
   For simplicity, we consider only the annealed case where 
the contribution of the charged monomers 
to the free energy becomes
\begin{eqnarray}
  \beta F_{\rm pa}^{el} = -\int\dr 
    \phi^2(\rr) \log\left(\sum_j p_j\e^{-\beta z_j e\psi(\rr)}\right)
  \label{betaFelpa}
\end{eqnarray}
and should be compared to the last term in eq.~\ref{betaFa}. 
The modified PB equation (eq.~\ref{PBa}) 
is now: 
\begin{eqnarray}
       \label{paPBa}
       \nabla^2\psi(\rr)
   &=& \frac{8\pi e}{\varepsilon}c_{\rm b} \sinh(\beta e\psi)
    -  \frac{4\pi e}{\varepsilon}
       \left( z_{\rm a}\phi^2 - \bar{z}\phi_{\rm b}^2 \e^{\beta e\psi} \right)
\end{eqnarray}
   where $\bar{z}\equiv \sum_j p_j z_j$ is the average monomer 
charge (in units of $e$) and
\begin{eqnarray}
    z_{\rm a}(\rr)={ \sum_j p_j z_j \e^{-\beta z_j e\psi(\rr)} \over
          \sum_j p_j     \e^{-\beta z_j e\psi(\rr)} }
\end{eqnarray} 
is the annealed (weighted) monomer valence, which is nothing but
the Boltzmann average of the monomer charge distribution.
   Finally, the SCF equation (eq.~\ref{SCFa}) is:
\begin{eqnarray}
   \label{paSCFa}
   \asix \nabla^2\phi(\rr) &=& v(\phi^3-\phi_{\rm b}^2\phi) 
       -\phi \log\left( \sum_j p_j\e^{-\beta z_j e\psi}\right)
\end{eqnarray}

\section{Discussion of the Charge Models}

In the previous section we derived the free energies and
mean-field equations for different charge distributions, 
by path integral methods.  

\bigskip
\noindent
{\bf The smeared case:~~~}The smeared free energy 
can be separated into uncharged polymer and
Coulombic contributions \cite{epl}, $F_{\rm s}=F_{\rm pol}+F_{\rm el}$.
The polymer part is:

\begin{eqnarray}
   F_{\rm pol} = k_B T \int\dr \Biggl\{
      &&~ \asix|\nabla\phi|^2 + \half v\phi^4(\rr)
     -\mu_{\rm p}\phi^2(\rr)
     \Biggr\}
  \label{betaFspol}
\end{eqnarray}
This is the Edwards free energy for polymer solutions
expressed in terms of the polymer order parameter $\phi(\rr)$, 
which is the square root of 
the monomer concentration, 
$\rho_m=\phi^2$. The first term represents the stiffness of the
polymer chains, the second takes into account
the excluded volume in good solvent conditions, while the
last term represents the coupling to a reservoir
of monomers with chemical potential $\mu_{\rm p}$. 

The Coulombic free energy reads:

\begin{eqnarray}
   F_{\rm el} = \int\dr \Biggl\{
    & - & \frac{\eps}{8\pi}|\nabla\psi|^2
     + k_B T\sum_\pm c_{\rm b}^\pm (1-\e^{\mp \beta e\psi(\rr)}) 
      +   p e\psi(\rr)\phi^2(\rr)
     \Biggr\}
  \label{betaFsel}
\end{eqnarray}
 This free energy
contains the electrostatic interactions between  all charges (small ions
and charged monomers), as well as the translational entropy
of the small ions in solution. 
In the above expression the  
independent fields are the electric
potential $\psi$ and polymer order parameter $\phi$.
The small ion concentrations are uniquely determined by the electric
potential through the Boltzmann weight:
$c^\pm(\rr)=c_{\rm b}^\pm\exp(\mp\beta e \psi(\rr))$.

The same free energy $F_{\rm el}$ can be obtained
from a more direct approach as will be shown for the smeared
case. It is convenient to express the free energy in terms
of the total charge density: $\rho_c=ec^{+}-ec^{-}+pe\phi^2$

\begin{eqnarray}
\tilde{F}_{\rm el}&=& {1\over 2}
\int \dr\drr{{\rho_c(\rr)\rho_c(\rr')}\over{\varepsilon |\rr-\rr'|}} \nonumber \\ 
&+&k_B T\int\dr\bigl\{c^{+}(\log c^{+}-1)+c^{-}(\log c^{-}-1)\bigr\}
-\int\dr\bigl\{\mu_{+}c^{+}+\mu_{-}c^{-}\bigr\}
\label{Felrho}
\end{eqnarray}

The free energy ${F}_{\rm el}$ of eq.~\ref{betaFsel}
should be distinguished from $\tilde{F}_{\rm el}$.
The former depends on $\psi$ and $\phi$ while the latter 
depends on $c^\pm$ and $\phi$. For the latter, 
the electric potential 
can be defined as
\begin{eqnarray}
\psi(\rr)=\int\drr{\rho_c(\rr')\over \varepsilon|\rr-\rr'|} 
\end{eqnarray}
and it satisfies the Poisson equation 
\begin{eqnarray}
\nabla^2\psi=-{4\pi \over \varepsilon}\rho_c
\label{Poisson}
\end{eqnarray}

Minimizing $\tilde{F}_{\rm el}$ with respect to
$c^\pm$ we obtain the equilibrium charge distribution
of the small ions:

\begin{eqnarray}
c^\pm(\rr)=c_{\rm b}^\pm\exp(\mp\beta e \psi(\rr)
\end{eqnarray}
where $\mu_\pm=k_B T \log c_{\rm b}^\pm$.

Substituting the above equilibrium condition back into eq.~\ref{Felrho}
we obtain
\begin{eqnarray}
\tilde{F}_{\rm el}= {1 \over 2}\int \dr {pe\phi^2 \psi}
-{1\over 2}\int \dr{(ec^{+}-ec^{-})\psi} \nonumber \\
-k_B T \int \dr\bigl\{c_{\rm b}^{+}(\e^{-\beta e \psi}-1)+
c_{\rm b}^{-}(\e^{\beta e \psi}-1)\bigr\}
\end{eqnarray}
From the Poisson equation \ref{Poisson}
we can express $c^{+}-c^{-}$ 
in terms of  $\nabla^2 \psi$ and $\phi$. Integration by parts
of the term $\psi\nabla^2\psi$ yields exactly the
first term of eq.~\ref{betaFsel} with the {\it correct}
negative sign \cite{epl,LL}.

\bigskip
\noindent
{\bf The annealed case:~~~}The 
second type of charge distribution is the annealed one where each
monomer can be either charged or neutral with bare probabilities
$p$ and $1-p$, respectively.
Its free energy eq.~\ref{betaFa} 
is similar to the smeared one except for the
coupling term between the charged monomers and the local electric
potential.  The difference can be understood in the following way:
in the annealed case different 
charge configurations will contribute to the free energy and
one needs to trace over those configurations in the partition function
{\it before} the free energy is calculated. In the smeared
case there is only one charge configuration where every monomer
is assigned a fractional charge $p e$, whereas in the annealed case
$p$ represents the {\it bare} probability of dissociation and
$p_a=p\exp(-\beta e\psi)/(1-p+p\exp(-\beta e\psi))$ is the 
{\it effective} probability as can be seen in eqs. \ref{PBa} and \ref{SCFa}.

In experiments the effective probability $p_a$ is related 
to the pH of the solution via
\begin{eqnarray}
{\rm pH}={\rm pK}_0 + \log_{10} {p_a\over 1-p_a}
\label{pH}
\end{eqnarray}
where ${\rm pK}_0=-\log_{10}{\rm K}_0$ and $K_0$ is the dissociation
constant. For example, in the case of weak alkaline monomers
\begin{eqnarray}
{\rm AOH} \rightleftharpoons {\rm A^{+}~+~OH^{-}}
\end{eqnarray}
K$_0$ is given by
\begin{eqnarray}
{\rm K_0={[A^{+}][OH^{-}]\over [AOH]}}
\end{eqnarray}
From eq.~\ref{pH} one can easily obtain $p_a$ as a 
function of the pH. 

At low electrostatic potentials $|\beta e\psi|\ll 1$, 
the annealed free energy can be expanded in powers
of $\psi$. The first term is equal to the smeared free energy 
$F_{\rm s}$,
while the next term is always negative

\begin{eqnarray}
  \beta F_{\rm a} \simeq \beta F_{\rm s}
     - \half  p(1-p)  \beta^2 e^2\int\dr \psi^2(\rr)\phi^2(\rr)
               ~~ < ~~\beta F_{\rm s}
 \label{betaFa2}
\end{eqnarray}
The fact that $F_{\rm a} < F_{\rm s}$ is related
to the convexity of the free
energy.  Indeed, the annealed 
case has more degrees of freedom and allows a better 
minimization.

\bigskip
\noindent
{\bf The permuted case:~~~}The 
permuted model is a  variant of the annealed case.
It models either mobile charges which can hop along the chain
or charges which associate and dissociate while keeping
the total amount of charge fixed on each chain.
 
The free energy for the permuted case can be written as the 
smeared free energy eq.~\ref{betaFs} plus an additional term of
entropic origin:
\begin{eqnarray}
  F_{\rm p} = F_{\rm s}
        + k_B T \int\dr 
\bigl\{ p\log p ~ + ~(1-p)\log(1-p) \bigr\} \phi^2(\rr)
  \label{Fp}
\end{eqnarray}
The correction is due to the translational entropy
of the charges along the chains. It amounts to a shift in the
polymer chemical potential and thus does not affect
the mean field equation.
  
  The correspondence  between the permuted and the smeared models
was not emphasized in previous works. It
can be interpreted as a tendency of the charges in the 
permuted model to spread uniformly along the linear chain.
However, as was discussed
earlier \cite{Raphael}, changing the pH of the solution (e.g., by
titration) can lead to non trivial dependence of $\mu_p$
 on the physical
parameters, since the reservoir concentration
changes in a titration process and will affect $\mu_{\rm p}$.

\bigskip
\noindent
{\bf The quenched case:~~~} In order to obtain the equilibrium
state of chains with frozen (quenched) 
charge distributions, the free energy
has to be averaged over all possible charge configurations. 
As was shown in the previous section, annealed and 
quenched polyelectrolytes
in contact with an infinite reservoir of chains have the same
mean field free energy.

The physical meaning of this result can be explained
in the following way \cite{Cates}: when quenched polymers 
are allowed to exchange with a bulk reservoir, 
containing all possible configurations, 
the system picks up  the optimal configurations from the bulk.
When the polymers are not coupled to an infinite reservoir
the annealed and quenched cases are different.
Note that the dynamics 
of annealed and quenched polyelectrolytes 
can differ considerably, 
but this is beyond the scope of the current work. 

\bigskip
\noindent
{\bf The annealed polyampholyte case:~~~}This 
situation corresponds to monomers which can 
carry a positive or negative charges 
of valency $z_j=0, \pm 1, \pm 2, \dots $ with probability $p_j$.
As was derived in the previous section the electrostatic
part of the free energy is given by:
\begin{eqnarray}
  F_{\rm pa}^{el} = -k_B T\int\dr 
    \phi^2(\rr) \log\left(\sum_j p_j\e^{-\beta z_j e\psi(\rr)}\right)
  \label{Fpa}
\end{eqnarray}

For low electrostatic potentials, 
an expansion of the above polyampholyte free energy yields:
\begin{equation}
   F_{\rm pa}^{el} \simeq 
  \bar{z} e\int\dr\psi(\rr)\phi^2(\rr)
     - \half \beta  \sigma^2 e^2 \int\dr \psi^2(\rr)\phi^2(\rr)    
  \label{betaFpa}
\end{equation}
The first term  is linear  in $\psi$ 
and reduces to the smeared contribution where $\bar{z}e$ replaces $pe$
as the average monomer charge. The second term
is a negative correction (as expected) which depends on the  
statistical variance of the charge distribution
\begin{eqnarray}
  \label{sigma2}
     \sigma^2=\sum_j p_j z_j^2-\Bigl(\sum_j p_j z_j\Bigr)^2
\end{eqnarray}
This term affects  the monomer-monomer correlations 
and $S(q)$ as discussed
in the following section.

The general distribution treated here has a few simple and useful
limits: 
\begin{enumerate}
\item
The smeared model (Sec.~III.A) is exactly recovered
for $p_1=1$ and $z_1=p$ yielding $\bar{z}=p$ and $\sigma^2=0$.

\item
The annealed model (Sec.~III.B) is obtained
for $p_1=p$, $z_1=1$ and $p_2=1-p$, $z_2=0$;
 $\bar{z}$ being simply $p$, $z_{\rm a}$ being $p_{\rm a}$,
and the variance $\sigma^2=p(1-p)$.

\item
A trimodal charge distribution can be a good representation
for  some polyampholyte systems.
\begin{equation}
f(q)=p_{+}\delta(q-e)+p_{0}\delta(q)+p_{-}\delta(q+e)
\end{equation}
where $p_0=1-p_{+}-p_{-}$. Each monomer can be either positively
charged, negatively charged or neutral with probabilities $p_{+}$, 
$p_{-}$ and $p_{0}$, respectively. The trimodal distribution
is characterized by two independent parameters. These can be
the first two moments:
the average $\bar{z}=p_{+}-p_{-}$ and the 
variance (``width") $\sigma^2=
p_{+}(1-p_{+})+p_{-}(1-p_{-})+2p_{+}p_{-}$. Increasing
the weight of $p_{0}$ means that the polyampholyte becomes
weakly charged, while increasing the weight of $p_{+}$ (or
$p_{-}$) means that the polyampholyte becomes more asymmetric
and resembles more a true polyelectrolyte.

\item 
A bimodal distribution of polyampholytes where each monomer carries
either a $+e$ charge or $-e$ one, with probabilities
 $p$ and $1-p$, respectively. 
\begin{equation}
f(q)=p\delta(q-e)+(1-p)\delta(q+e)
\end{equation}
with $\bar{z}=2p-1$ and 
 $\sigma^2=4p(1-p)$.
Note that for 
this fully charged polyampholyte
the same parameter $p$ characterizes {\it both}
the average $\bar{z}$ and the variance $\sigma^2$, 
so that $\sigma^2=1-\bar{z}^2$. The distribution is symmetric 
around $p=1/2$. As $|p-1/2|$ increases, $|\bar{z}|$ increases,
$\sigma^2$ decreases, and the polyampholyte
 resembles more and more a polyelectrolyte with a net charge.
\end{enumerate}

It is of interest to consider explicitly the symmetric bimodal 
case
mentioned above having no net charge
$p=1/2$, $\bar{z}=0$ and $\sigma^2=1$ is maximal.
The PB equation (eq.~\ref{paPBa}) can be written as
\begin{eqnarray}
       \label{paPBsym}
       \nabla^2\psi(\rr)
   &=& \frac{8\pi e}{\varepsilon}c_{\rm b} \sinh(\beta e\psi)
    +  \frac{4\pi e}{\varepsilon}\phi^2\tanh(\beta e\psi)
\end{eqnarray}
   and the SCF equation (eq.~\ref{paSCFa}) is:
\begin{eqnarray}
   \label{paSCFsym}
   \asix \nabla^2\phi(\rr) &=& v(\phi^3-\phi_{\rm b}^2\phi) 
       -\phi \log\left[ \cosh(\beta e\psi)\right]
\end{eqnarray}
The second term in eq.~\ref{paPBsym} represents the 
contribution of the charged monomers to the local charge density. 
At low potentials $|\beta e\psi|\ll 1$ the polymer charge density 
is $e\phi^2\tanh(\beta e\psi) \simeq e\phi^2\sinh(\beta e\psi)$.
Comparing, in this limit, the two terms on the right hand side of
eq. \ref{paPBsym}, 
the polyampholyte can be viewed as a
symmetric electrolyte \cite{Dobrynin_PA} 
whose bulk concentration is not a constant
but determined by the local monomer concentration.
For a monovalent
 electrolyte, the local concentrations of positive 
and negative ions obey a Boltzmann distribution 
 $c^\pm(\rr) = c_{\rm b}\exp(\mp\beta e\psi)$, where 
the equilibrium distribution is achieved by exchange of ions 
with the reservoir.
For the annealed polyampholytes, the concentrations of 
positive and negative monomers $\rho_{\rm m}^\pm(\rr)$ 
behave in a similar way (for weak potentials),  
 $\rho_{\rm m}^\pm(\rr) \simeq \half\phi^2\exp(\mp\beta e\psi)$, but 
the mechanism is different. For the latter,
the equilibrium distribution 
is achieved by ionizing the monomers, and this process is 
limited by the monomer concentration, 
whereas for the former case, 
the reservoir contains an infinite amount of ions.
This difference becomes evident at 
high potentials where $\tanh(\beta e\psi)$ saturates to $+1$ or 
 $-1$, depending on the sign of the potential. 
Under these extreme conditions the polyampholytes are no longer
neutral. Instead, they are fully ionized.

\section{Structure Factor within RPA}

   Density--density correlations are measured in scattering experiments
\cite{Candau,Schmidt}
and can be calculated using the random phase approximation (RPA)
\cite{Erukhimovich,LeiblerJoanny,Raphael,Vilgis,Dormidontova,Liu}.
This is done by considering small fluctuations of the homogeneous bulk
state. Since we are interested in the monomer-monomer density
correlations, we do not perform the integration over $\rho_{\rm m}(\rr)$ in 
eq.~\ref{Zvarphi}, and express the partition function in terms 
of the three fields $\rho_{\rm m}(\rr) = \phi_{\rm b}^2 + \delta \rho_{\rm m}(\rr)$,
 $\varphi_{\rm m}(\rr) = \varphi_{\rm m}^{(0)} + \delta \varphi_{\rm m} (\rr)$ and 
 $\varphi_{\rm c}(\rr) = \varphi_{\rm c}^{(0)} + \delta \varphi_{\rm c} (\rr)$.
As introduced in Sec.~II, $\varphi_{\rm m}$ is the conjugate field of the local
monomer concentration $\rho_{\rm m}$ and   $\varphi_{\rm c}$ is the
electric potential conjugate to the charge density $\rho_{\rm c}$.  
Note that $\varphi_{\rm c}^{(0)}=0$ in the bulk.

The free energy is then expanded to second order 
in these fluctuations.
Since the linear terms in $\delta \rho_{\rm m}$, 
$\delta \varphi_{\rm m}$ and $\delta \varphi_{\rm c}$
cancel out,
the first non-zero corrections are of second order.

It is more convenient
to write this expansion 
in Fourier space:
\begin{equation}
\delta\rho_{\rm m}(\rr) =\int \dq \delta\rho_{\rm m}(\q)\,\e^{i\q\cdot\rr}
\end{equation}
and similarly for $\delta\varphi_{\rm m}$ and $\delta\varphi_{\rm c}$.

\subsection{Smeared $S(q)$}

Using the smeared free energy, the expansion 
 yields
\begin{eqnarray}
  \beta F_{\rm s}(\delta\rho_{\rm m},\delta\varphi_{\rm m},\delta\varphi_{\rm c}) &\simeq& 
      \int \dq \Biggl\{ 
        v \delta \rho_{\rm m} (-\q) \delta \rho_{\rm m} (\q)  
      + v^2 \phi_{\rm b}^2 S_0(q) \delta\varphi_{\rm m}({-\q})\delta\varphi_{\rm m}({\q})
       \nonumber \\ && 
      + \beta {\eps\over 4\pi} 
            \Bigl[ q^2 + \kappa_s^2 + p\kappa_p^2 S_0 (q) \Bigr]
            \delta\varphi_{\rm c}({-\q})\delta\varphi_{\rm c}({\q})
        \nonumber \\ && 
      + \beta p e v \phi_{\rm b}^2 S_0(q) \delta\varphi_{\rm m}(-\q) \delta\varphi_{\rm c}(\q)
      - i v \delta\rho_{\rm m} (-\q) \delta\varphi_{\rm m}(\q)
       \Biggr\}  
  \label{fRPAs}
\end{eqnarray}
where the integral is over the wavevector ${\bf q}$, and
$S_0(q)$ is the structure factor of Gaussian chains.
For chains of length $N$, it is equal to
$S_0(q)=N D({1\over 6}a^2q^2 N)$ 
where $D(\zeta)=(2/\zeta^2)(\e^{-\zeta}+\zeta-1)$
is the Debye function \cite{deGennesBook}.
For infinitely long chains $N\rightarrow\infty$, the structure factor
is independent of the chain length, $S_0^{-1}(q)={1\over 12}a^2q^2$.

The Debye--H\"uckel screening length 
has two contributions:
\begin{eqnarray}
\kappa_{\rm tot}^2=\kappa_s^2+p\kappa_p^2=8\pi l_Bc_{\rm b}+
 4\pi l_B p \phi_{\rm b}^2
\end{eqnarray}
 the first contribution arises
from the 1:1 symmetric electrolyte  while
the second one comes from the polymer counterions,
 $l_B=e^2/\eps\kt$ being the 
Bjerrum length,  equal to about 
$7$\AA\ at room temperature. 

The Fourier transform of the monomer--monomer correlations
 $\langle\delta\rho_{\rm m}(\rr)\delta\rho_{\rm m}({\bf 0})\rangle$
is proportional to the experimentally measured structure factor $S(q)$. It
can now be calculated 
as a Gaussian integral giving:
\begin{eqnarray}
   S_{\rm s}^{-1}(q) = 
   \frac{\phi_{\rm b}^2}{ \langle\delta\rho_{\rm m}(-\q)\delta\rho_{\rm m}(\q)\rangle} =
  S_0^{-1}(q) + v\phi_{\rm b}^2 + 
  {p^2\kappa_p^2\over q^2+\kappa_s^2+p\kappa_p^2}
  \label{RPAs}
\end{eqnarray}
 Similar expressions were obtained in previous works
\cite{Erukhimovich,LeiblerJoanny,Raphael,Vilgis,Dormidontova},
using somewhat different derivations.

\subsection{Annealed  $S(q)$ for 
Polyelectrolytes and Polyampholytes}

Repeating the above calculation for the 
{\it annealed} case leads to the same form for $S(q)$
where $\kappa_{\rm tot}^2$ is replaced 
by $\kappa_{\rm tot}^2 + p(1-p)\kappa_p^2$ in eq. \ref{RPAs}.
\begin{eqnarray}
   S_{\rm a}^{-1}(q) = 
  S_0^{-1}(q) + v\phi_{\rm b}^2 + 
  {p^2\kappa_p^2\over q^2+\kappa_s^2+p(2-p)\kappa_p^2}
  \label{RPAa}
\end{eqnarray}
The enhanced 
screening here 
is due to the additional annealed degrees of freedom 
of the charges on the polymer chains.
Local fluctuations of the monomer charge density
effectively increase the local ion concentration
leading to stronger screening.

The structure factor
is also calculated  for annealed polyampholytes.
It amounts to replacing
$\kappa_{\rm tot}^2$ 
by $\kappa_s^2+(|\bar{z}|+\sigma^2)\kappa_p^2$
and $p^2\kappa_p^2$ by $\bar{z}^2\kappa_p^2$
in eq. \ref{RPAs}
leading to:
\begin{eqnarray}
   S_{\rm pa}^{-1}(q) = S_0^{-1}(q) + v\phi_{\rm b}^2 
             + {\bar{z}^2\kappa_p^2 \over
               q^2+\kappa_s^2+(|\bar{z}|+\sigma^2)\kappa_p^2}
  \label{RPApa}
\end{eqnarray}
 For neutral (symmetric) polyampholytes,
$\bar{z}=0$, expression \ref{RPApa} is the same as $S^{-1}(q)$
of neutral polymers. We note that this is an outcome of the RPA 
which neglects higher order charge correlations. However, the mean field
equations themselves as well as the free energy \ref{Fpa} depend on the
charge distribution and, in particular, on the variance $\sigma^2$.

 At high charge fraction $|\bar{z}|$ and low salt concentration, 
the structure factor $S(q)$ exhibits a peak at a 
finite wavenumber $q_0>0$ satisfying 
\begin{eqnarray}
   \left[ q_0^2+\kappa_s^2+(|\bar{z}|+\sigma^2)\kappa_p^2 \right]^2 
   = 12\bar{z}^2\kappa_p^2/a^2
   \label{q0_PA}
\end{eqnarray}
This peak is characteristic of polyelectrolyte solutions at low
salt concentration.  Since $S^{-1}(q)$ is the energy of the $q$
mode
density fluctuations, a maximum in $S(q)$ corresponds to the 
lowest energy fluctuation. Here $q_0>0$ results from the competition
between the first term in \ref{RPApa}, originating from the polymer
elasticity, and the last term due to the electrostatic
interactions screened by the small ions. For $q$ values
$\kappa <q< a^{-1}$, the peak $q_0$ can be observed.
For example, in the absence of salt, the condition 
for having a peak 
at $q_0>0$ is 
\begin{eqnarray}
0< {\sigma^2\over |\bar{z}|} < {2\sqrt{3}\over \kappa_p a}-1
\label{Peak}
\end{eqnarray}
The upper bound is bigger than zero as long as 
$\kappa_p^{-1}\ge a$ 
recalling that 
$\kappa_p^{-1}=1/\sqrt{4\pi l_B \phi_{\rm b}^2}$ and $a$ 
is the monomer size.
The above 
inequality is satisfied for highly asymmetric polyampholytes,
resembling polyelectrolytes
(see case 3 of the preceding section). 
In the opposite limit, low $|\bar{z}|$ and/or high $\sigma$, 
the random
polyampholyte behaves essentially as a neutral
polymer ($q_0\rightarrow 0$). 
For fully charged polyampholytes (case 4 of
the  preceding section),
$\sigma^2/|\bar{z}|=(1-\bar{z}^2)/|\bar{z}|$ yielding the same 
conclusion. As long as the net charge $|\bar{z}|$ is large
enough, the polyampholytes resemble polyelectrolytes and
$S(q)$ will exhibit a peak at finite $q_0$.

In Fig.~2 
the structure factor $S(q)$ is plotted as function of 
the  wavenumber $q$ for smeared polyelectrolytes
($\sigma^2=0$) for different salt concentrations and charge
fractions $\bar{z}$. 
The structure factor at wavenumber $q=0$  is proportional 
to the osmotic compressibility.
As depicted in Fig.~2, $S(q=0)$ increases upon 
addition of salt, while the peak position at $q_0$ shifts  
to smaller wavenumbers until the peak
disappears and 
 $S(q)$ becomes a monotonous decreasing 
function.
   The inset of Fig.~2 shows the effect of decreasing the 
average monomer charge $|\bar{z}|$. The peak 
increases and shifts towards smaller values of $q$.

\subsection{Mesophases in Bad Solvent}

We end this section 
by examining  polyelectrolytes and
polyampholytes
in bad solvent conditions.
The excluded volume parameter $v$ is negative leading to 
collapse (and segregation)  of the chains, and higher order
virial terms have to be included in the free energy. Assuming that the 
third order virial coefficient ${w}$ is positive, 
with a contribution of ${1\over 6}w \rho_{\rm m}^3$ to the free energy,
we obtain 
\begin{eqnarray}
   S_{\rm pa}^{-1}(q) = S_0^{-1}(q) - |v|\phi_{\rm b}^2 + 
{w}\phi_{\rm b}^4 
             + {\bar{z}^2\kappa_p^2 \over
               q^2+\kappa_s^2+(|\bar{z}|+\sigma^2)\kappa_p^2}
  \label{RPApab}
\end{eqnarray}
An analogous expression was obtained for smeared polyelectrolytes 
\cite{Erukhimovich,LeiblerJoanny,Dormidontova}
and is generalized here to annealed polyampholytes.

An instability of the homogeneous phase
is determined by $S^{-1}(q_0)=0$, indicating
 a {\it mesophase} separation, where 
the size of the micro--domains is $\lambda_0=2\pi/q_0$.
In Fig.~3, the line marking the instability
of the homogeneous (disordered) phase is plotted
for different salt concentrations.
At high ionic
strength dilute polyelectrolyte solutions becomes unstable. 
This reflects the {\it macroscopic} phase separation
 of neutral (screened) polymers in a bad solvent 
($q_0\rightarrow 0$ as is seen in Fig.~2).

In Fig.~4 we show the 
dependence of the micro-domain region
on the variance of the charge distribution, $\sigma^2$.
One important consequence of eq.~\ref{RPApab} is
the equivalence of electrostatic screening induced by
fluctuations in the polyampholyte 
charges and 
screening by added salt with effective bulk
concentration 
 $c_{\rm b}^{\rm eff}={1\over 2} \sigma^2\phi_{\rm b}^2$.
For example, for $\phi_{\rm b}^2=10^{-6}$\AA$^{-3}$, 
 $c_{\rm b}^{\rm eff}/\sigma^2 = 0.8$mM. 
This effect is stronger in
polyampholytes than in annealed polyelectrolytes,
since in almost symmetric polyampholytes
the excess charge is small, $|\bar{z}|\ll 1$,
 while the variance of the charges $\sigma^2$ can be 
close to one.

It is instructive to look at the 
instability of the homogeneous system towards macro-phase 
separation as a function of the polyampholyte average net charge.
We emphasize that in some cases this macro-phase separation
occurring at $q=0$ is preempted by a mesophase
at $q_0>0$. Nevertheless, let us consider the change in the 
second virial coefficient $v\sim T - T_{\theta}^0$
due to the electrostatic interactions
where $T_{\theta}^0$ is the $\theta$-temperature
in the absence of electrostatic interactions. We note that for 
 $\bar{z}=0$ there is no electrostatic contribution
to $S(q)$ in our RPA calculation. For $\bar{z}\ne 0$, the $q=0$
instability will occur for
\begin{equation}
|v|\phi_{\rm b}^2=
{\bar{z}^2\kappa_p^2\over \kappa_s^2+(|\bar{z}|+\sigma^2)\kappa_p^2}
\end{equation}

For fully charged polyampholytes (taking the bimodal
distribution: $z_j=\pm 1$ and $|\bar{z}|\le 1$),
the $\theta$-temperature in
the  presence of charged monomers is
\begin{equation}
T_{\theta}^0- T_{\theta}(\bar{z}) \sim
{\bar{z}^2\over 2c_{\rm b}/\phi_{\rm b}^2 + |\bar{z}|+1 -\bar{z}^2}
\label{DeltaTtheta}
\end{equation}
This behavior is similar to the one found for single
chains \cite{Kantor}. In Fig.~5 we plot
the dependence of the $\theta$-temperature on the net charge 
 $\bar{z}$  for different salt 
concentrations.
As expected, the $\theta$-temperature is a decreasing
function of the net charge $|\bar{z}|$. 
%
%
As can be seen from Fig.~5, addition of salt extends the 
bad solvent regime 
to higher values of $|\bar{z}|$.
At high salt concentrations, the electrostatic interactions 
are screened and  $T_{\theta}(\bar{z})$ is very close to its
pure value $T_{\theta}^0$.

\section{Conclusions}
\label{conclusions}

   In this work we have studied bulk properties of
charged polymers in aqueous solutions in the presence of added salt.
Starting from a path integral formalism which takes into account
the chain connectivity, short range and electrostatic interactions
we derived
 mean--field equations describing
the behavior of polyelectrolytes and polyampholytes
in solution. We  compared several models for the 
statistical charge distribution corresponding to
different experimental realizations.
The simplest and 
most frequently used model is the smeared one where
charges are uniformly distributed.
The permuted model, where the charges are mobile 
along the chain, is found to be equivalent to the smeared model
except for a constant shift in the monomer chemical potential.
This shift has to be taken into account
in titration experiments.

The annealed model was found to have a lower free energy than
the smeared one. This is related to the additional degrees of freedom.
The effective dissociation depends on the local electric
potential.
At thermodynamic equilibrium, 
the quenched case
is found to be  
equivalent to the  annealed one, 
as long as the system is in contact
with an infinite reservoir (bulk) 
of polyelectrolyte chains.

Annealed polyampholytes are characterized by their net charge
and variance. We find different behavior
for symmetric polyampholytes (no net charge) as compared
to asymmetric ones (closer to polyelectrolytes). 
At low electrostatic potentials, all the above mentioned 
models have 
the same limiting behavior. 

The monomer--monomer 
structure factor $S(q)$ is calculated within 
the Random Phase Approximation for annealed 
polyelectrolytes and polyampholytes.
The electrostatic screening depends not only on
the salt and counterions but also on
the variance of the 
annealed charge distribution.

A peak in $S(q)$ for polyampholytes
at finite wavenumber is shown to
appear at a high  net charge and/or low variance
indicating polyelectrolyte-like behavior. 
For bad solvent conditions, the variance 
enhances the tendency of the system to undergo
a mesophase separation.

Finally, it is worth mentioning that 
the present 
study can be further extended to treat polyelectrolytes
and polyampholytes in restricted geometries and close to charged surfaces
\cite{FleerBook,Varoqui91,epl,macromols,Podgornik}.
In particular, it will be interesting to address the question
of how the different charge distribution is coupled with
the polymer adsorption
onto a single surface and 
the forces exerted by polyelectrolytes and polyampholytes
between planar, cylinders or spherical surfaces.

\vspace*{0.5cm}
{\em Acknowledgments}

   We would like to thank L. Auvray, N. Dan,
H. Diamant, J. Israelachvili,
 Y. Kantor
and S. Safran for useful discussions.  
   Two of us (IB and DA) would like to thank 
the Service de Physique Th\'eorique (CE-Saclay) 
and one of us (HO) 
the School of Physics and Astronomy (Tel Aviv University)
for their warm hospitality.
 Partial support from the Israel Science Foundation 
founded by the Israel Academy of Sciences and Humanities 
-- centers of Excellence Program 
and the U.S.-Israel Binational Foundation (B.S.F.) 
under grant No. 94-00291 gratefully acknowledged. 


\pagebreak
\section*{Figure Captions}
\pagestyle{empty}

\begin{description}

\item {\bf Fig.~1}:
  Schematic view of a polyelectrolyte solution. The monomer
coordinates are $\rr_l(s)$ where $l=1,...,M$ labels the polymer 
chain and $s\in[0,N]$ is a continuous index along the chain.
The small ion coordinates are $\Rp$ and $\Rm$ where 
 $i=1,...,N^+$ and $j=1,...,N^-$.

\item {\bf Fig.~2}:
The effect of salt concentration on the structure factor $S(q)$ 
of polyelectrolytes, eq.~\ref{RPApa}.
The parameters used are:
 $\phi_{\rm b}^2=10^{-6}$\AA$^{-3}$, $a=5$\AA, $v=0.1a^3$,
 $\bar{z}=0.1$ and $\sigma^2=0$.
The salt concentration is:
 $c_{\rm b}=1$mM (solid curve);
 $c_{\rm b}=2$mM (dots);
 $c_{\rm b}=5$mM (short dashes) and
 $c_{\rm b}=10$mM (long dashes).
The inset shows the effect of the average monomer charge 
at small salt concentration $c_{\rm b}=0.1$mM.
The different curves correspond to:
 $\bar{z}=0.10$ (solid curve);
 $\bar{z}=0.03$ (dots) and
 $\bar{z}=0.01$ (dashes).

\item {\bf Fig.~3}:
The effect of salt concentration on the stability line
of the homogeneous phase.
The parameters used are:
 $a=5$\AA, ${w}=5\times 10^5$\AA$^6$, $\bar{z}=0.01$ 
and $\sigma^2=0$.
The salt concentration is:
 $c_{\rm b}=0$ (solid curve);
 $c_{\rm b}=1$mM (short dashes);
 $c_{\rm b}=1.5$mM (dots and short dashes);
 $c_{\rm b}=2.5$mM (long dashes) and
 $c_{\rm b}=5$mM (dots and long dashes).

\item {\bf Fig.~4}:
The effect of the  variance $\sigma^2$ 
 on the stability line of the homogeneous phase.
The parameters used are:
 $a=5$\AA, ${w}=5\times 10^5$\AA$^6$, $\bar{z}=0.01$ 
and $c_{\rm b}=0.01$mM.
The different curves correspond to
 $\sigma^2=0$ (solid curve);
 $\sigma^2=0.01$ (short dashes) and
 $\sigma^2=0.04$ (long dashes).

\item {\bf Fig.~5}: 
The shift in the $\theta$-temperature of polyampholytes:
$\Delta T_{\theta}=T_{\theta}-T_{\theta}^0$ in units of 
$T_{\theta}^0$
as function of the excess charge $|\bar{z}|$ for different 
salt concentrations, eq.~\ref{DeltaTtheta}.
Only the fully charged polyampholyte case: $z_j=\pm 1$ is shown.
The polymer concentration is $\phi_{\rm b}^2=10^{-6}$\AA$^{-3}$,
$|v|\phi_{\rm b}^2=0.01$
and the salt concentration:
 $c_{\rm b}=0$ (solid curve);
 $c_{\rm b}=10$mM (dots);
 $c_{\rm b}=0.1$M (dashes);
 and $c_{\rm b}=1$M (long dashes).  
\end{description}

\pagebreak \vfill

\begin{figure}[tbh]
  {\Large Fig.~1}
  \bigskip\bigskip\bigskip

  \epsfxsize=0.5\linewidth
  \centerline{\hbox{ \epsffile{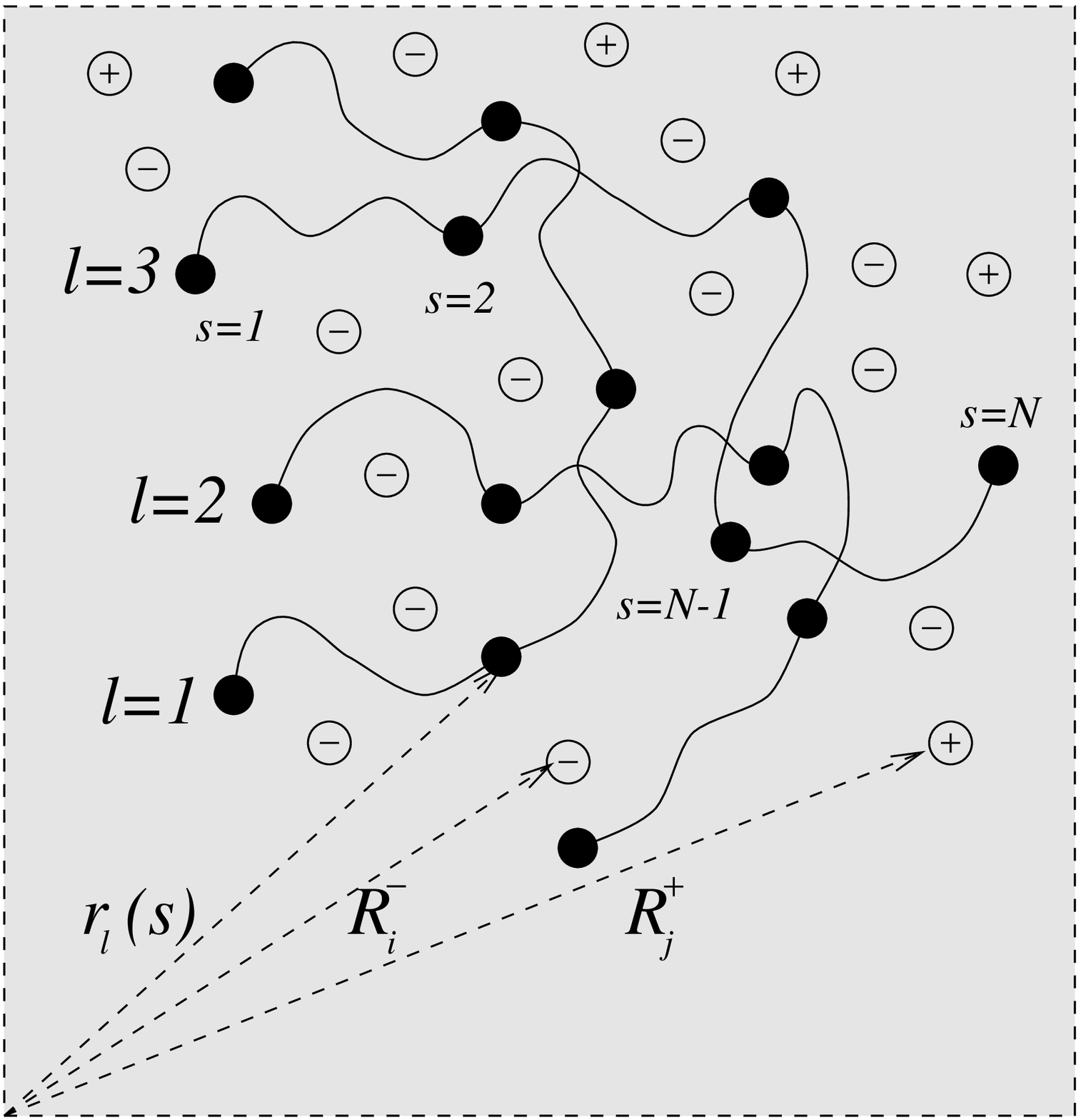} }}
\end{figure}
\vfill

\begin{figure}[tbh]
  {\Large Fig.~2}
  \bigskip\bigskip\bigskip

  \epsfxsize=0.5\linewidth
  \centerline{\hbox{ \epsffile{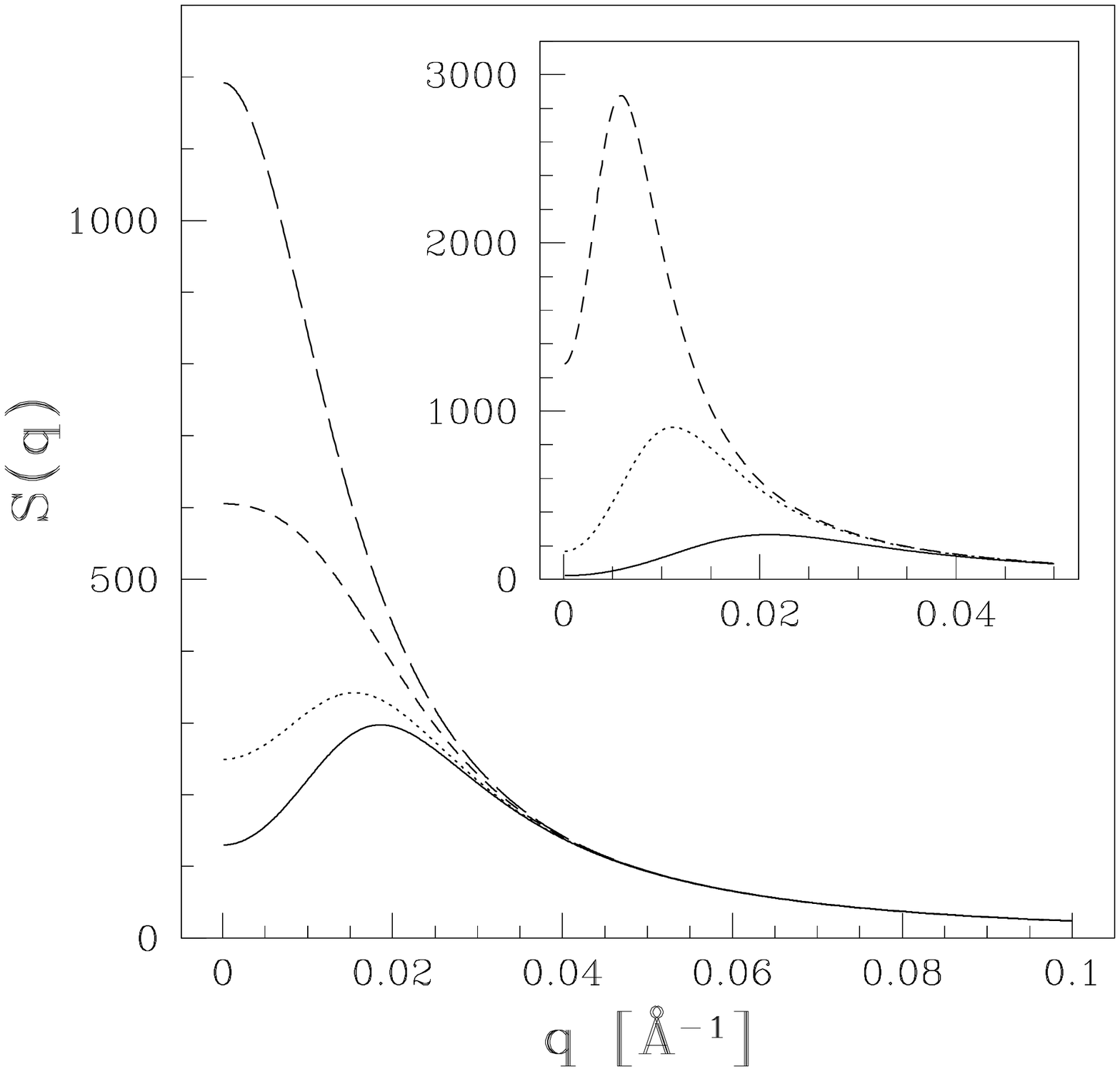} }}
\end{figure}
\vfill

\pagebreak \vfill

\begin{figure}[tbh]
  {\Large Fig.~3}
  \bigskip\bigskip\bigskip

  \epsfxsize=0.5\linewidth
  \centerline{\hbox{ \epsffile{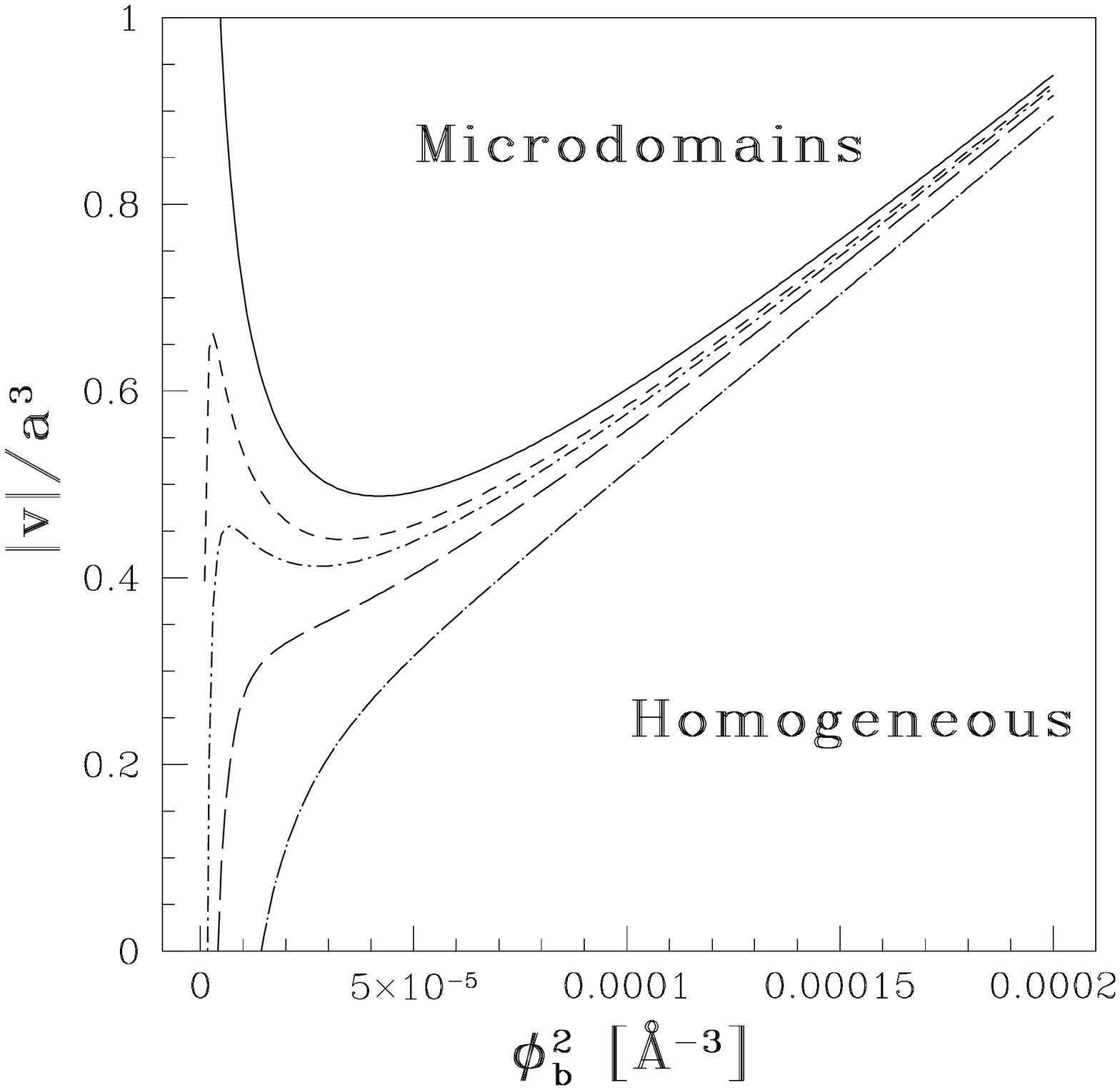} }}
\end{figure}
\vfill

\begin{figure}[tbh]
  {\Large Fig.~4}
  \bigskip\bigskip\bigskip

  \epsfxsize=0.5\linewidth
  \centerline{\hbox{ \epsffile{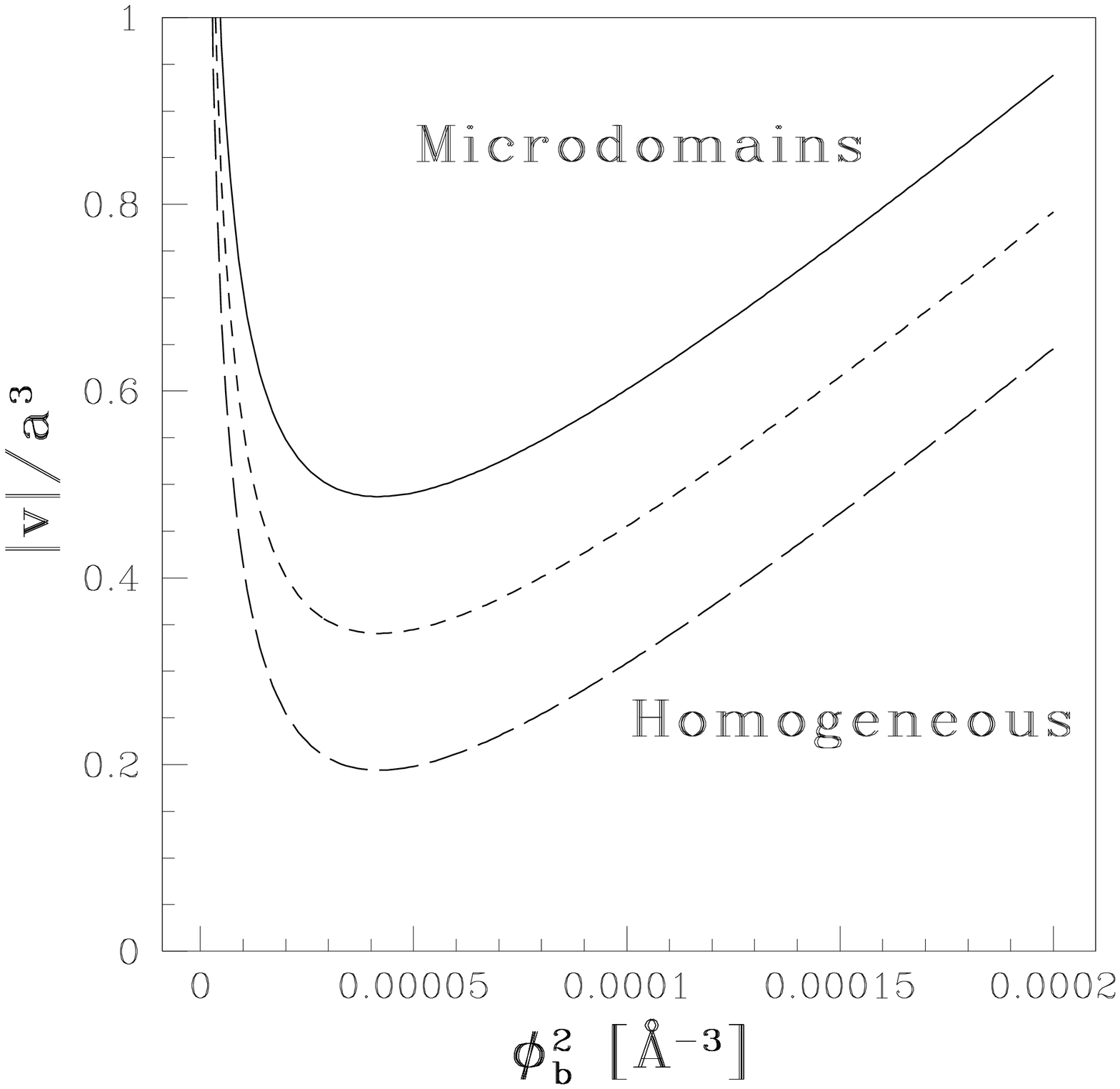} }}
\end{figure}
\vfill

\begin{figure}[tbh]
  {\Large Fig.~5}
  \bigskip\bigskip\bigskip

  \epsfxsize=0.5\linewidth
  \centerline{\hbox{ \epsffile{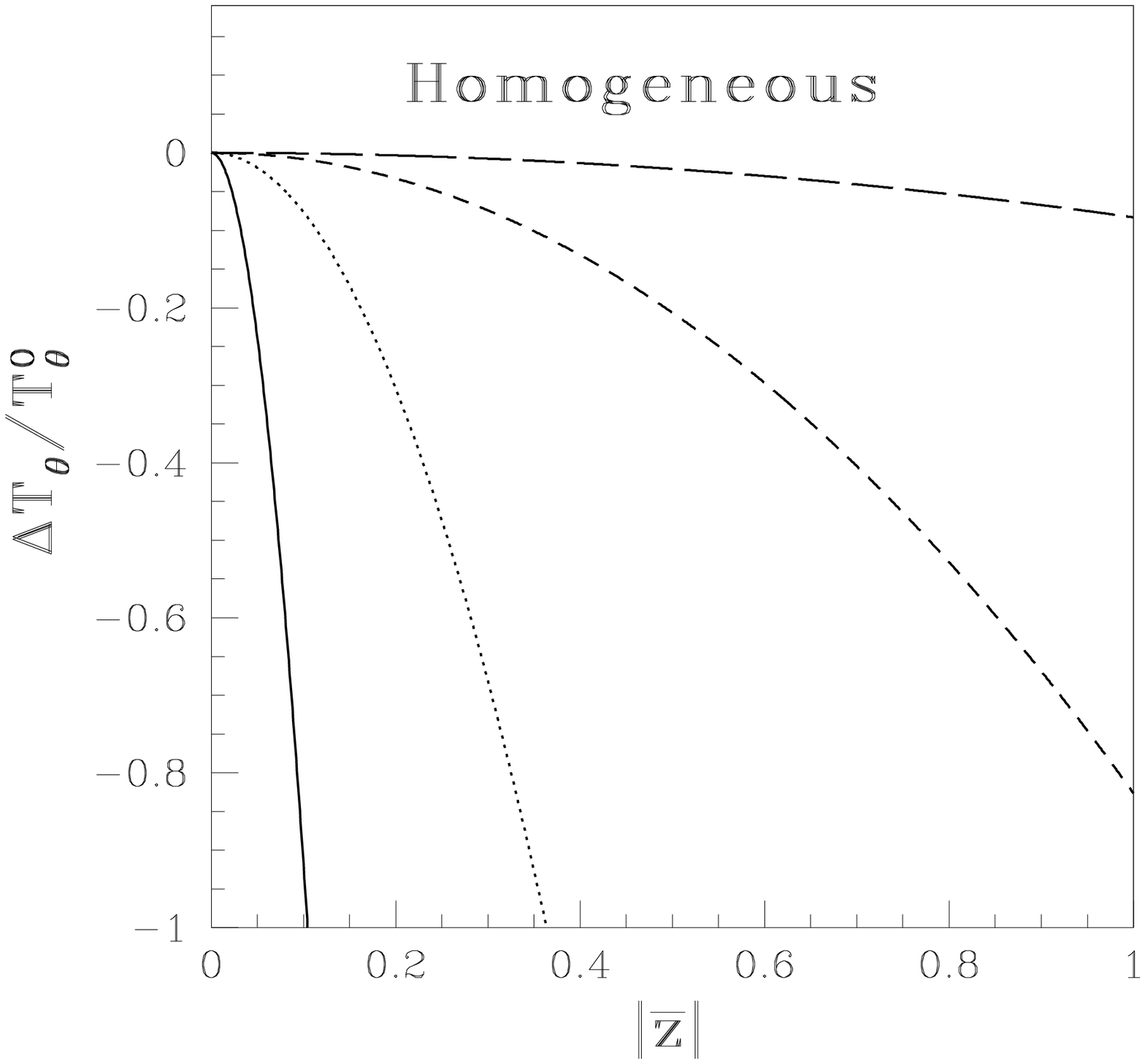} }}
\end{figure}
\vfill

\end{document}